\documentclass[twocolumn]{aastex61}

\defcitealias{2016ApJ...817..120C}{C16}
\usepackage{amsmath,amssymb}

\shorttitle{\emph{HST} followup of two bright $\MakeLowercase{z} \sim 8$ candidate galaxies from BoRG}
\shortauthors{Livermore et al.}
\begin{document}

\title{\emph{HST} followup observations of two bright $\MakeLowercase{z} \sim 8$ candidate galaxies from the B\MakeLowercase{o}RG pure-parallel survey}

\correspondingauthor{R.~C. Livermore}
\email{rlivermore@unimelb.edu.au}

\author[0000-0003-4456-1566]{R.~C. Livermore}
\affil{School of Physics, The University of Melbourne, VIC 3010, Australia}
\affiliation{ARC Centre of Excellence for All Sky Astrophysics in 3 Dimensions (ASTRO 3D)}

\author{M. Trenti}
\affiliation{School of Physics, The University of Melbourne, VIC 3010, Australia}
\affiliation{ARC Centre of Excellence for All Sky Astrophysics in 3 Dimensions (ASTRO 3D)}

\author{L.~D. Bradley}
\affiliation{Space Telescope Science Institute, 3700 San Martin Drive, Baltimore, MD 21218, USA}

\author{S.~R. Bernard}
\affiliation{School of Physics, The University of Melbourne, VIC 3010, Australia}

\author{B.~W. Holwerda}
\affiliation{Department of Physics and Astronomy, University of Louisville, Louisville, KY 40292, USA}

\author{C.~A. Mason}
\affiliation{Department of Physics and Astronomy, University of California, Los Angeles, CA 90095, USA}

\author{T. Treu}
\affiliation{Department of Physics and Astronomy, University of California, Los Angeles, CA 90095, USA}



\begin{abstract}
We present follow-up imaging of two bright ($L > L^{\ast}$) galaxy candidates at $z \gtrsim 8$ from the Brightest of Reionizing Galaxies (BoRG) survey with the F098M filter on the \emph{Hubble Space Telescope}/Wide Field Camera 3 (\emph{HST}/WFC3). The F098M filter provides an additional constraint on the flux blueward of the spectral break, and the observations are designed to discriminate between low- and high-$z$ photometric redshift solutions for these galaxies. Our results confirm one galaxy, BoRG\_0116+1425\_747, as a highly probable $z \sim 8$ source, but reveal that BoRG\_0116+1425\_630 - previously the brightest known $z > 8$ candidate ($m_{\rm{AB}} = 24.5$) - is likely to be a $z \sim 2$ interloper. As this source was substantially brighter than any other $z > 8$ candidate, removing it from the sample has a significant impact on the derived UV luminosity function in this epoch. We show that while previous BoRG results favored a shallow power-law decline in the bright end of the luminosity function prior to reionization, there is now no evidence for departure from a Schechter function form and therefore no evidence for a difference in galaxy formation processes before and after reionization.
\end{abstract}




\section{Introduction} \label{sec:intro}

The unprecedented sensitivity of the Wide Field Camera 3 (WFC3) on the \emph{Hubble Space Telescope} (\emph{HST}) has provided new insight into the build-up of galaxies during the epoch of reionization \citep[$z \gtrsim 7$;][]{2011ApJ...737...90B,2015ApJ...803...34B,2013MNRAS.432.2696M,2015ApJ...814...69AA,2015ApJ...810...71F,2015ApJ...799...12I,2017ApJ...835..113L}. The picture that has emerged from these observations is that the UV luminosity function is well described by a Schechter function at least out to $z \sim 7$ with the characteristic luminosity and normalization decreasing and the faint-end slope becoming steeper with increasing redshift \citep[e.g.][]{2016PASA...33...37F}.

Before the completion of reionization, though, the limited sample sizes mean that the picture is less clear. A key remaining question is whether the UV luminosity function continues to show an exponential cut-off at the bright end \citep[e.g.][]{2012ApJ...760..108B,2014ApJ...786...57S,2015ApJ...803...34B} or whether it can be described by a power law \citep[e.g.][]{2014MNRAS.440.2810B,2015MNRAS.452.1817B,2017MNRAS.466.3612B,2015ApJ...810...71F,2016ApJ...817..120C}.

Results from semi-analytic models and observations have shown that the break at the bright end of the luminosity function can be interpreted as evidence for quenching due to feedback from active galactic nuclei \citep[AGNs;][]{2006MNRAS.370..645B,2006MNRAS.365...11C} or some other mass-quenching law \citep{2010ApJ...721..193P}, or the build-up of dust in the brightest galaxies \citep{2014MNRAS.440.3714R}. Studies have shown that the impact of magnification bias, which might conceal an exponential break, is negligible for current surveys \citep{2015MNRAS.450.1224B,2015ApJ...805...79M}. Therefore, if the exponential break is not seen at higher redshifts, this might indicate that the feedback processes are less efficient prior to the completion of reionization, or that brighter galaxies have not yet built up the requisite dust content \citep[e.g.][]{2017arXiv171006628D}.

In addition to providing insight into the feedback processes affecting galaxy evolution, the UV luminosity function can also provide a view of the overall evolution of star formation in the universe. The integral of the UV luminosity function is related to the cosmic star formation rate density, which is known to increase with redshift up to around $z \sim 2$ and then decline \citep[e.g.][]{1996MNRAS.283.1388M,2014ARA&A..52..415M}. Studies of the decline beyond $z \sim 8$ have had conflicting results, with some recent work suggesting a smooth decline at $z > 8$ \citep{2013ApJ...763L...7E,2016MNRAS.459.3812M}, while others favor a transition to a much steeper decline \citep{2014ApJ...786..108O,2015ApJ...808..104O}. Resolving this tension is vital both for constraining models of galaxy evolution and for accurately predicting expected detections of high-redshift galaxies with the \emph{James Webb Space Telescope} (\emph{JWST}).

The Brightest of Reionizing Galaxies (BoRG) survey \citep[PI: M. Trenti;][]{2011ApJ...727L..39T} uses pure-parallel observations with \emph{HST}/WFC3 to obtain random pointings across the sky, resulting in coverage over a wide area that is designed to search for rare, bright objects at high redshift. The first results of the BoRG[z9-10] survey revealed five candidate galaxies at $8.3 < z < 10$, including the brightest known candidate at $z > 8$ \citep[hereafter \citetalias{2016ApJ...817..120C}]{2016ApJ...817..120C}. This galaxy exhibits a strong spectral break in the F105W filter, suggesting a redshift of $z > 8$. However, \citetalias{2016ApJ...817..120C} find a $\sim 10\%$ probability that this is instead a 4000\,\AA\ break at $z \sim 1.8$. If confirmed to be at $z > 8$, this brightest candidate with $M_{\rm{UV}} = -22.8$ (apparent $H$-band magnitude $m_{H160} = 24.5$) would provide strong evidence in favor of an excess of galaxies at the bright end of the UV luminosity function, lending support to a power-law decline as opposed to the exponential cut-off.

In this Letter, we present follow-up observations with \emph{HST} to confirm the high-redshift nature of this brightest candidate, which also cover a second source in the same field discovered by \citetalias{2016ApJ...817..120C}. We describe the observations and data reduction in Section \ref{sec:obs}, and re-derive the photometric redshift of the two high-$z$ candidates in Section \ref{sec:sources}. In Section \ref{sec:lf} we present the revised UV luminosity function at $z \sim 9$, and we summarize our conclusions in Section \ref{sec:conc}. Throughout, magnitudes are given according to the AB system \citep{1983ApJ...266..713O} and we adopt a \citet{2016A&A...594A..13P} $\Lambda$CDM cosmology.

\section{Observations and Data Reduction}
\label{sec:obs}

\begin{figure*}[htb!]
\plotone{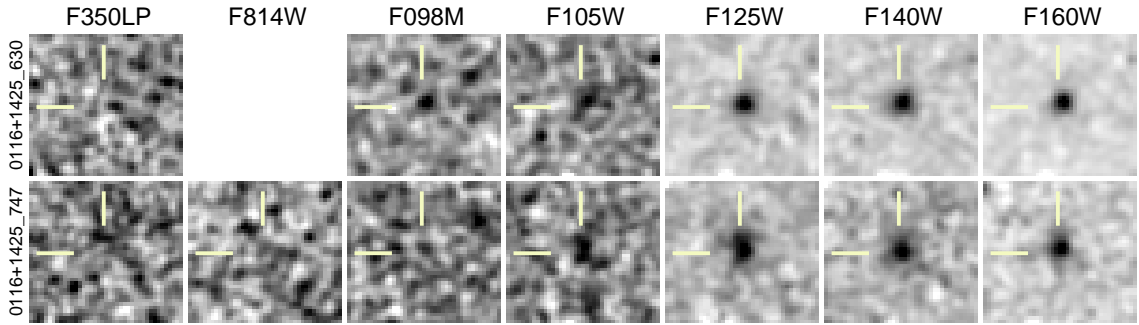}
\caption{Postage stamp images of the two high-$z$ candidates, BoRG\_0116+1425\_630 (upper) and BoRG\_0116+1425\_747 (lower) in each filter. Each image is centered on the source and measures $3\farcs 2 \times 3\farcs 2$. Note that there is no F814W coverage of BoRG\_0116+1425\_630.}
\label{fig:stamps}
\end{figure*}

\begin{deluxetable*}{l r r C C C C C C C r}
\tabletypesize{\footnotesize}
\tablewidth{\textwidth}
\tablecaption{High-$z$ candidates in the BoRG\_0116+1425 field}
\label{tab:mags}
\tablecolumns{11}
\tablehead{ \colhead{ID} & \colhead{$\alpha$(J2000)} & \colhead{$\delta$(J2000)} & \multicolumn{7}{c}{AB Magnitude\tablenotemark{a}} & \colhead{$z$} \\
\colhead{} & \colhead{(degree)} & \colhead{(degree)} & \colhead{F350LP} & \colhead{F814W} & \colhead{F098M} & \colhead{F105W} & \colhead{F125W} & \colhead{F140W} & \colhead{F160W} & \colhead{} }
\startdata
BoRG\_0116+1425\_630 & 19.0347 & 14.4026 & > 28.58\tablenotemark{b} & \nodata & 26.79 \pm 0.33 & 26.88 \pm 0.34 & 25.38 \pm 0.08 & 25.08 \pm 0.07 & 24.81 \pm 0.06 & 1.8 \\
BoRG\_0116+1425\_747 & 19.0372 & 14.4068 & > 28.78\tablenotemark{c} & > 28.46\tablenotemark{d} & > 28.21\tablenotemark{e} & 27.00 \pm 0.32 & 25.97 \pm 0.11 & 25.98 \pm 0.13 & 25.70 \pm 0.10 & 7.9 \\
\enddata
\tablenotetext{a}{Magnitudes are isophotal fluxes from {\sc SExtractor} FLUX\_ISO. For reference, the total magnitudes in F160W for the two sources (from MAG\_AUTO) are $24.48 \pm 0.06$ and $25.04 \pm 0.10$ for 0116+1425\_630 and 0116+1425\_747 respectively. Measured fluxes for values given as upper limits are: $^b$\,$4 \pm 13$\,nJy; $^c$\, $10 \pm 11$\,nJy; $^d$\,$10 \pm 15$\,nJy; $^e$\,$18 \pm 19$\,nJy.} 
\tablecomments{Coordinates and magnitudes of the two high-$z$ candidates discovered by \citetalias{2016ApJ...817..120C} in the BoRG\_0116+1425 field. Columns 2-3 give the $\alpha$ and $\delta$ coordinates in degrees. Columns 4-10 give the magnitude (from {\sc SExtractor} FLUX\_ISO) in each band (note there is no F814W coverage of BoRG\_0116+1425\_630). For non-detections, we quote the 1$\sigma$ uncertainty as an upper limit. The redshift $z$ in column 11 is the photometric redshift obtained from BPZ as described in the text.}
\end{deluxetable*}

The design of the BoRG[$z9-10$] survey (Program ID: 13767; PI: M. Trenti) is described in \citetalias{2016ApJ...817..120C}. Briefly, it comprises 480 orbits of pure-parallel imaging of independent lines of sight with the near-IR WFC3 filters F105W, F125W, F140W, and F160W, as well as the long-pass optical filter F350LP. The large area and medium depth (5$\sigma$ point-source sensitivity of $m_{\rm{AB}} \sim 26.5 - 27.5$) is designed to constrain the bright end of the UV luminosity function at $z > 8$ by identifying bright galaxy candidates through broadband photometry.

In order to follow-up the bright $z > 8$ candidate BoRG\_0116+1425\_630 discovered by \citetalias{2016ApJ...817..120C}, an additional orbit was acquired in Cycle 24 (Program ID: 14701; PI: M. Trenti) with the F098M filter.

In the same field as BoRG\_0116+1425\_630 is another $z > 8$ candidate, BoRG\_0116+1425\_747, which has additional archival \emph{HST}/Advanced Camera for Surveys (ACS) coverage in the F814W filter (Program ID: 14652; PI: B. W. Holwerda). 

The calibrated data were downloaded from the Mikulski Archive for Space Telescopes (MAST), where individual exposures had been processed through the standard calibration software to apply bias correction (ACS and WFC3/UVIS), dark subtraction, flat-fielding, and charge-transfer efficiency (CTE) correction (ACS and WFC3/UVIS). As with previous BoRG analyses (\citealt{2012ApJ...760..108B,2014ApJ...786...57S,2016ApJ...827...76B}; \citetalias{2016ApJ...817..120C}), we applied Laplacian edge filtering \citep{2001PASP..113.1420V} to the WFC3 data to remove residual cosmic rays and detector artifacts such as unflagged hot pixels. Additionally, we corrected the WFC3/UVIS F350LP data for electronic crosstalk \citep{2012wfc..rept....2S}. The individual exposures in each filter were combined using {\sc AstroDrizzle} to produce the final science images and their associated inverse-variance weight maps. Like previous BoRG analyses \citep[e.g.][]{2011ApJ...727L..39T,2012ApJ...760..108B,2014ApJ...786...57S,2016ApJ...827...76B,2016ApJ...817..120C}, the images were drizzled to a final pixel scale of 0\farcs 08\,/pixel. The total exposures times of the combined images were 2095\,s, 5207\,s, 2612\,s, 2209\,s, 2059\,s, 1759\,s, 2409\,s in the F350LP, F814W, F098M, F105W, F125W, F140W, and F160W filters, respectively.

One limitation of pure-parallel observations is that they are not dithered, in order to avoid conflict with the primary observation. Accordingly, the design of the BoRG[$z9-10$] survey is optimized to mitigate the impact of the lack of dithering as far as possible. \citetalias{2016ApJ...817..120C} give a full list of the steps taken (their Section 2) and a comparison between undithered pure-parallel data and overlapping dithered data showing that the impact on photometry is negligible (their \S 3.1). We also note that star/galaxy separation is reliable down to approximately a magnitude above the photometric limit \citep{2014ApJ...788...77H}.

For the new data (F814W and F098M), we derived variance maps (rms) from the inverse-variance weight maps (wht) as $\rm{rms} = 1 / \sqrt{\rm{wht}}$. Slight correlation between pixels results in the weight maps underestimating the rms, so we rescale them by measuring photometry in empty apertures and normalizing the entire image by a constant factor such that the median error on the fluxes of empty apertures matches the variance of the sky flux measurements \citep[see][]{2011ApJ...727L..39T}. The normalization factors were 1.24 in F814W and 1.10 in F098M; for the pre-existing images analyzed by \citetalias{2016ApJ...817..120C} they range from 1.06 in F160W to 1.33 in F350LP. These noise measurements also allow us to derive $5\sigma$ limiting magnitudes in these two filters of $m_{\rm{AB}} = 26.60$ and $26.37$, respectively (for limiting magnitudes in the other filters, see \citetalias{2016ApJ...817..120C}, Table 1).

\begin{figure*}[htb]
\includegraphics[width=0.49\textwidth]{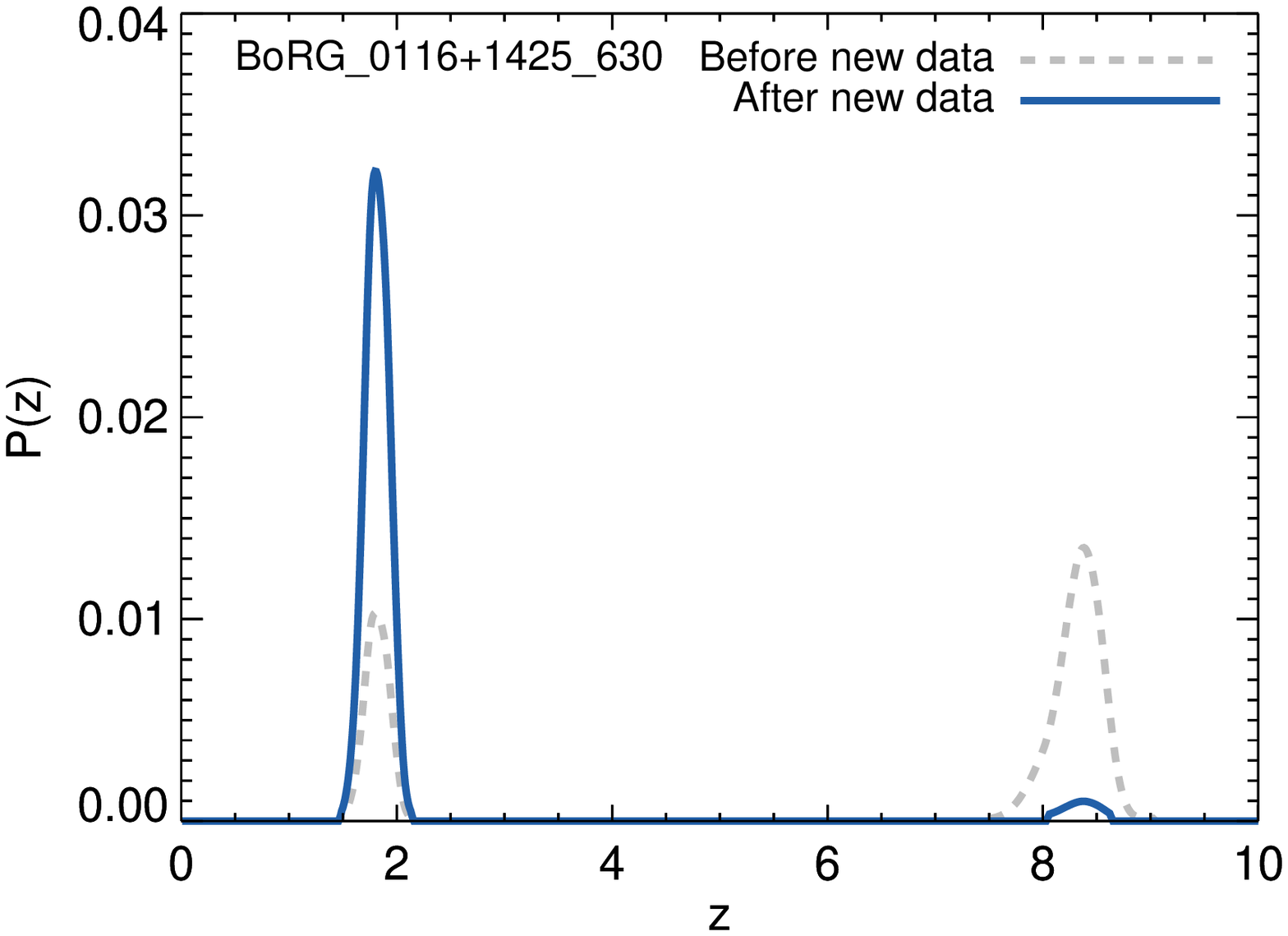}\hfill \includegraphics[width=0.49\textwidth]{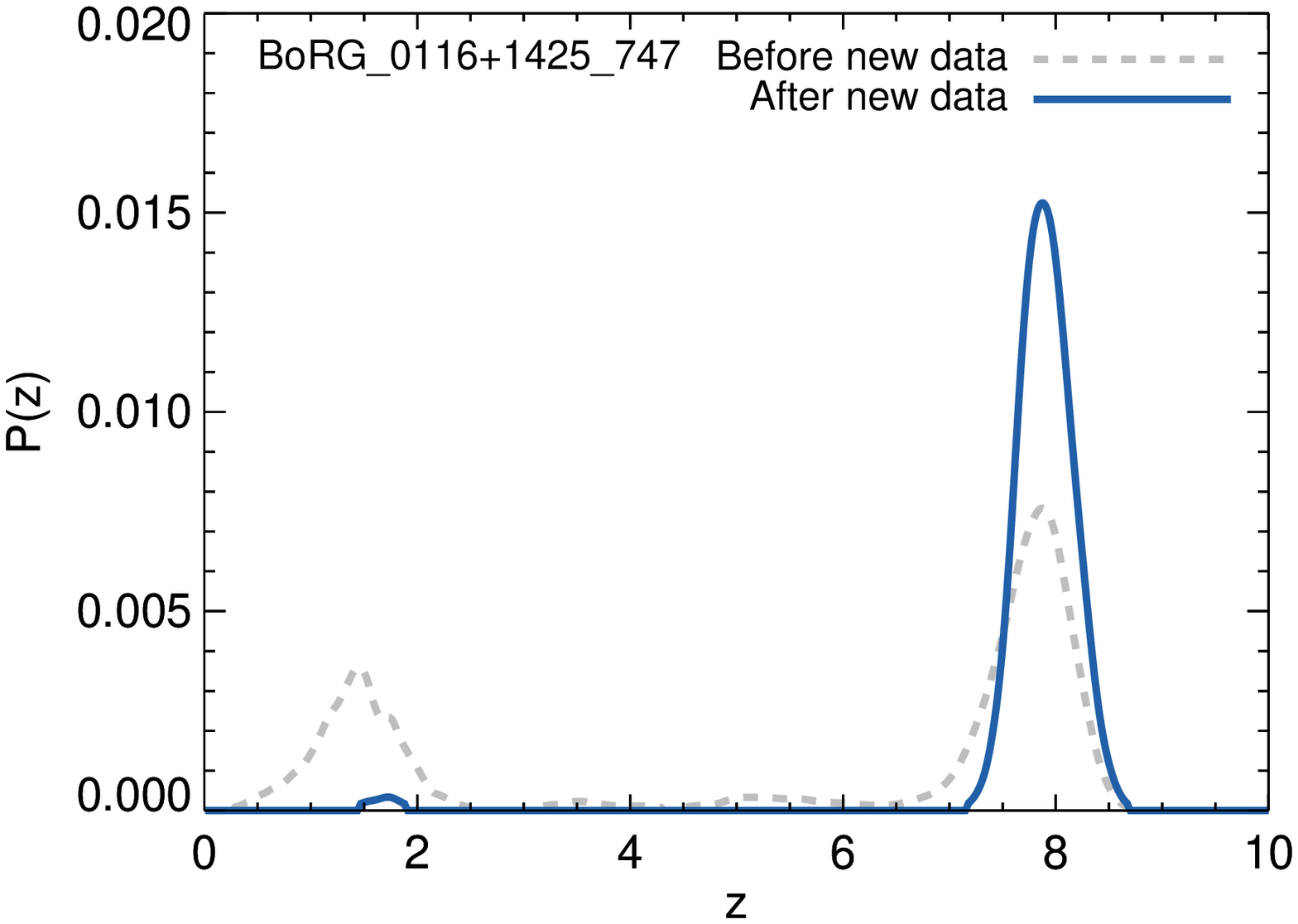}
\caption{Photometric redshift probability distribution ($P(z)$) obtained from BPZ with a flat prior. The gray dashed line indicates the $P(z)$ obtained before adding the new data, indicating the degeneracy of the $z \gtrsim 8$ and $z < 2$ solutions. The solid blue line shows the result after adding the F098M (and, for BoRG\_0116+1425\_747, F814W) imaging. The new addition of the new data causes the $P(z)$ to converge on one solution; a lower redshift $z \sim 1.8$ is now preferred for BoRG\_0116+1425\_630, whereas BoRG\_0116+1425\_747 remains likely to be at $z \sim 8$ with a higher probability.}
\label{fig:pz}
\end{figure*}

\begin{figure*}[htb]
\includegraphics[width=0.49\textwidth]{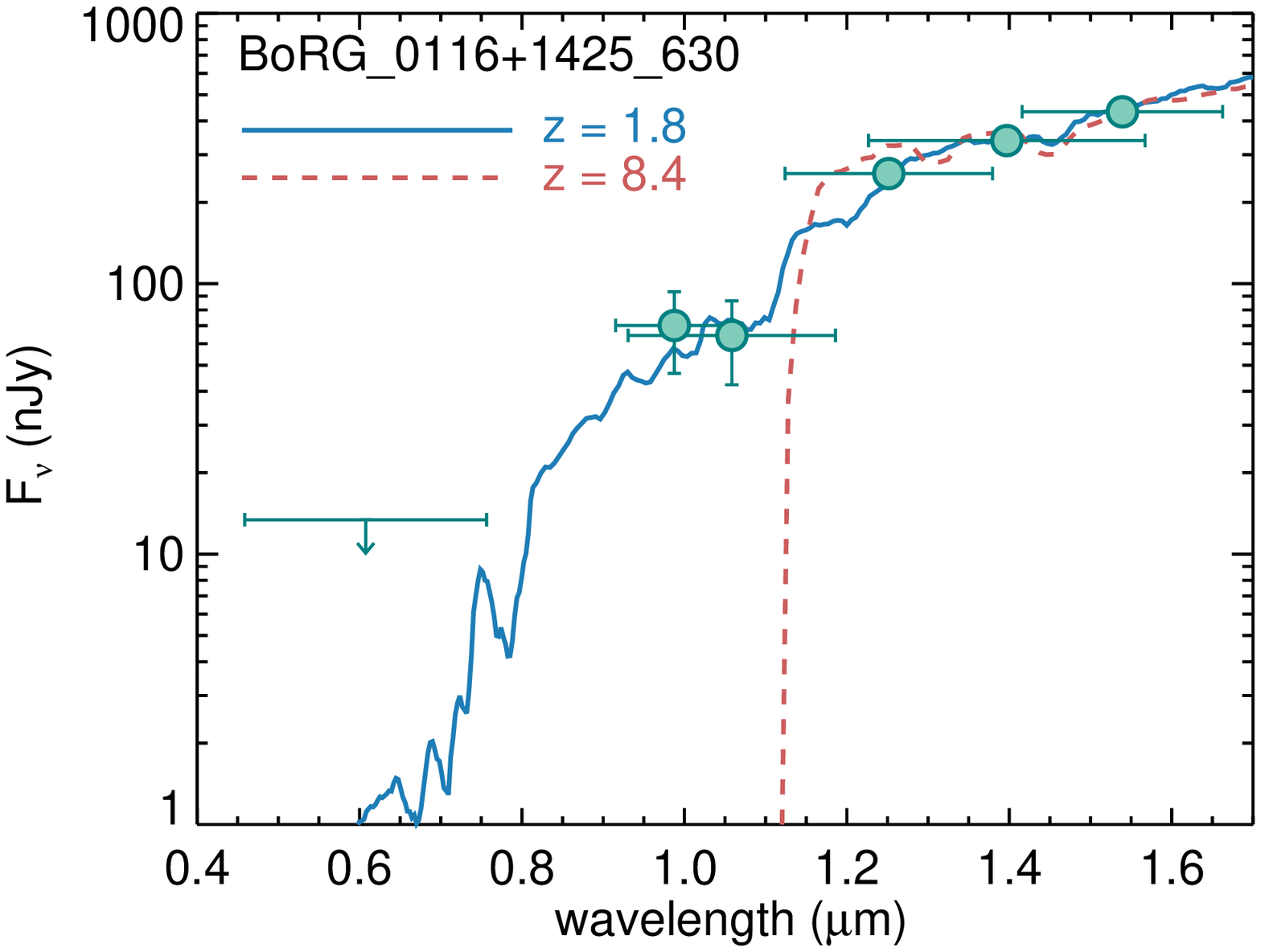}\hfill \includegraphics[width=0.49\textwidth]{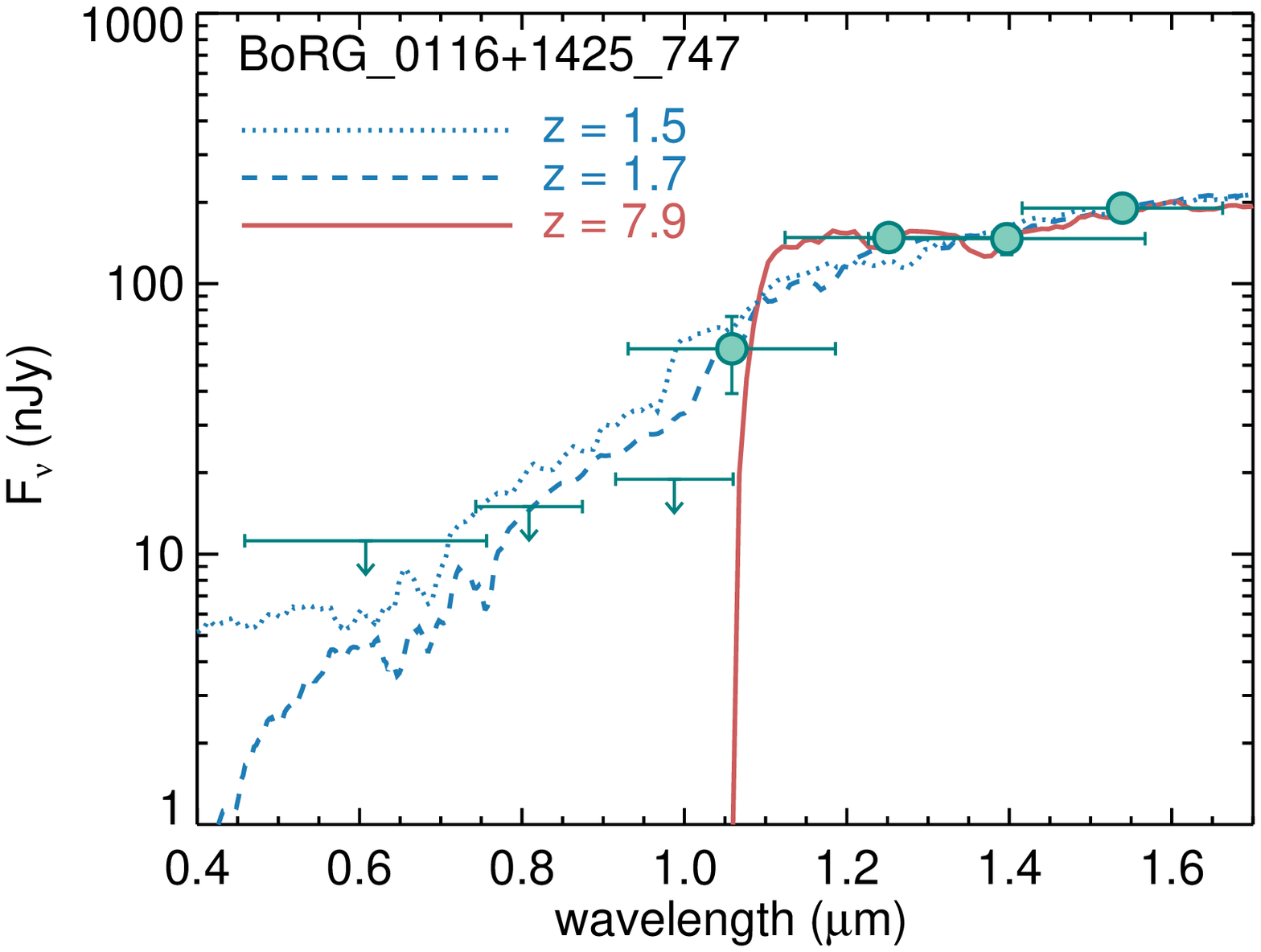}
\caption{Best-fit spectral energy distributions (SEDs) for the two sources at each of the peaks in the photometric redshift probability distribution. For BoRG\_0116+1425\_747 we show both the low-$z$ solution before adding the new data ($z \sim 1.5$) and that from after adding the new data ($z \sim 1.7$). In all other cases, the SED fits are identical before and after adding the new data. The F098M filter acts as a key differentiator between the low- (blue) and high- (red) redshift solutions. The SED preferred after adding the new data is indicated by a solid line.}
\label{fig:sed}
\end{figure*}

\section{High-\emph{\MakeLowercase{z}} Candidates}
\label{sec:sources}

We constructed source catalogs using {\sc Source Extractor} ({\sc SExtractor}; \citealp{1996AAS..117..393B}) in dual image mode, using a combined F140W and F160W image and corresponding combined weight map as the detection image. The {\sc SExtractor} parameters were chosen to match those of \citetalias{2016ApJ...817..120C}, so we require nine contiguous pixels with signal-to-noise ratio (S/N)$ > 0.7$. We corrected the fluxes for foreground Galactic extinction $E(B - V) = 0.0349$ \citep{2011ApJ...737..103S}, and following previous BoRG work \citep[see][]{2012ApJ...746...55T} we adopt the isophotal flux (FLUX\_ISO) for measuring colors, and define the signal-to-noise as $\rm{S/N} = \rm{FLUX\_ISO} / \rm{FLUXERR\_ISO}$. From this final catalog, we select the two high-$z$ candidates from \citetalias{2016ApJ...817..120C} and confirm that the same magnitudes are derived in all of the pre-existing filters. Postage stamp images of the two sources, BoRG\_0116+1425\_630 and BoRG\_0116+1425\_747 are shown in Figure \ref{fig:stamps}, and the positions and magnitudes are given in Table \ref{tab:mags}.

\citetalias{2016ApJ...817..120C} rely on a Lyman-break selection  to identify high-$z$ candidate galaxies, but it is nonetheless instructive to use photometric redshift codes to visualize the probability distribution of the redshift of the sources. We use the {\sc BPZ} code \citep{2000ApJ...536..571B,2006AJ....132..926C} both with and without the new F098M (and, for BoRG\_0116+1425\_747, F814W) data, and the results are shown in Figure \ref{fig:pz}. Both candidates show a break in the F105W filter, which, if interpreted as the Lyman break, places them at redshift $z \gtrsim 8$. In both cases, a secondary solution exists at $z < 2$, in the case where the break is instead the 4000\,\AA~break. As shown in Figures \ref{fig:pz} and \ref{fig:sed}, the addition of the F098M filter helps to distinguish between these two solutions as flux can be measured blueward of the shallower 4000\,\AA~break, but not the steeper Lyman break.

Figures \ref{fig:pz} and \ref{fig:sed} suggest that BoRG\_0116+1425\_630 is more likely to be a $z < 2$ contaminant, due to a $3\,\sigma$ detection in the F098M filter; the integral of the $P(z)$ contained in the low-$z$ solution has increased from 31\% prior to the addition of F098M, to 96\% afterwards. Meanwhile, BoRG\_0116+1425\_747, with signal-to-noise S/N$ < 1$ in F098M, now has a stronger probability (99\%, compared to 64\% previously) of being a $z \sim 8$ galaxy (an independent analysis with another photometric redshift code confirms the $z \sim 8$ solution in the latter case; Bridge et al. 2018, in preparation). With the addition of the new data, the secondary solutions encompass a negligible fraction of the probability distribution: 4\% for BoRG\_0116+1425\_630 and 1\% for BoRG\_0116+1425\_747. The best-fitting photometric redshifts are now $1.80^{+0.15}_{-0.10}$ and $7.87^{+0.29}_{-0.23}$ for BoRG\_0116+1425\_630 and BoRG\_0116+1425\_747 respectively, where the uncertainties include the central 68\% of the probability distribution. We note that adding the new data does not change the redshift of the high-$z$ peak of the redshift probability distribution in either case (to within the $\Delta z = 0.1$ resolution used in the fit). The low-$z$ peak for BoRG\_0116+1425\_630 is also unchanged, but the secondary solution for BoRG\_0116+1425\_747 has increased from $z = 1.45$ to $z = 1.73$.

We can also modify the Lyman-break selection method used by \citetalias{2016ApJ...817..120C} to incorporate the new data. These criteria require strong (S/N$ > 4$) detections in each filter redward of the Lyman break, with a relatively flat spectrum in this region ($JH_{140} - H_{160} < 0.3$), and a clear break in a pair of adjacent filters ($Y_{105} - JH_{140} > 1.5$ and $Y_{105} - JH_{140} > 5.33 \cdot \left(JH_{140} - H_{160}\right) + 0.7$), with a non-detection (S/N$ < 1.5$) blueward of the break. A cut of S/N$ \geq 8$ is also required in the detection image. Both of the candidates were previously selected to meet these criteria (see \citetalias{2016ApJ...817..120C} for full details). To this, we can add a requirement for S/N$ < 1.5$ in the F098M filter, as well as F814W for BoRG\_0116+1425\_747. With S/N$ \sim 3$ in F098M, BoRG\_0116+1425\_630 does not meet the new criteria and is therefore removed from the $z \sim 9$ sample. However, BoRG\_0116+1425\_747 has S/N$ < 1$ in both F814W and F098M, and so would still be selected in the $z \sim 8$ sample.

Using the available photometry, we can derive some physical characteristics of the two galaxies. We use the Fitting and Assessment of Synthetic Templates (FAST) code \citep{2009ApJ...700..221K} with \citet{2003MNRAS.344.1000B} stellar population synthesis models, a \citet{2003PASP..115..763C} initial mass function, a \citet{2000ApJ...533..682C} dust attenuation law and a delayed exponential star formation history. Assuming that BoRG\_0116+1425\_630 has the best-fitting redshift of $z \sim 1.8$, it is undetected in the rest-frame ultraviolet, giving an upper limit on the star formation rate of $\log (\rm{SFR}/ M_{\sun}\,yr^{-1}) < 0.47$, and we find a stellar mass $\log (M_{\ast}/M_{\sun}) = 9.93^{+0.28}_{-0.25}$. For BoRG\_0116+1425\_747, assuming $z \sim 7.9$, we find $M_{\rm{UV}} = -22.0$ and a star formation rate $\log (\rm{SFR}/ M_{\sun}\,yr^{-1}) = 1.13^{+1.28}_{-0.52}$.

\section{The UV Luminosity Function at $\MakeLowercase{z} \sim 9$}
\label{sec:lf}

\begin{figure}
\includegraphics[width=0.48\textwidth]{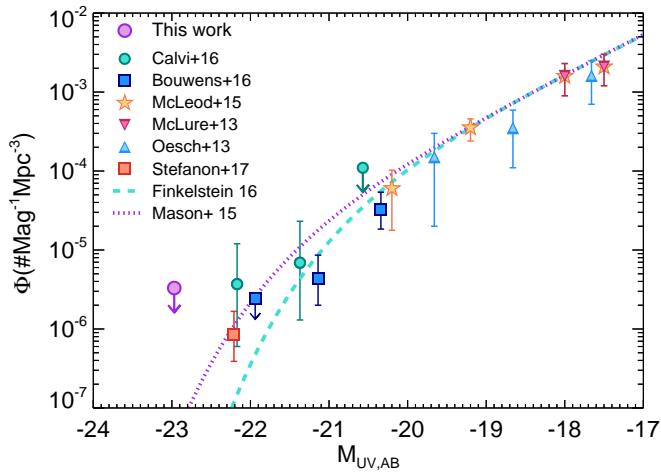}
\label{fig:lf}
\caption{The $z \sim 9$ luminosity function after the low-$z$ contaminant BoRG\_0116+1425\_630 is removed. Removing this bright source means that the excess at the bright end presented by \citetalias{2016ApJ...817..120C} becomes an upper limit, which is now consistent with the theoretical prediction of \citet{2015ApJ...813...21M}. Additional observational data are shown from \cite{2013MNRAS.432.2696M}, \cite{2013ApJ...773...75O}, \citet{2016ApJ...830...67B}, \citet{2016MNRAS.459.3812M}, and \citet{2017ApJ...851...43S}.}
\end{figure}

The new data presented in this Letter indicate that the $z \sim 9$ candidate BoRG\_0116+1425\_630 is a low-$z$ interloper. Due to its bright nature, the removal of this single source from the $z\sim 9$ BoRG sample has a significant effect on the derived luminosity function. With a derived absolute magnitude at $z \sim 9$ of $M_{\rm{AB}} = -22.8$, it would be extremely rare if the UV luminosity function were to remain Schechter-like in this epoch with an exponential cut-off at the bright end. Based on the theoretical prediction of the evolution of the UV luminosity function of \citet{2015ApJ...813...21M}, the probability of finding a galaxy at this magnitude in the BoRG data would be $p = 2.8\times 10^{-3}$. According to the empirically derived luminosity function of \citet{2016PASA...33...37F}, it would be even rarer, with a probability of $p = 5.4\times 10^{-5}$. Therefore, had this candidate been confirmed, it would therefore have been strong evidence in favor of a departure from a Schechter function in this epoch.

As we have shown that this galaxy is likely to be a $z \sim 2$ interloper, we revise the luminosity function to exclude this candidate. Full details of the computation of the luminosity function are provided in \citetalias{2016ApJ...817..120C}; in Figure \ref{fig:lf} we show the result of removing this single source. There are now no known sources at $z > 8$ with $M_{\rm{AB}} < -22.5$, and the observations are now consistent with the predicted Schechter function of \citet{2015ApJ...813...21M}. This means that there is no evidence that the process of galaxy evolution differs before and after reionization. However, we note that while the BoRG data now do not favor a power-law form to the bright end of the luminosity function, nor do they rule it out. There remains tentative evidence that a power law is preferred at $z \sim 7$ \citep{2017MNRAS.466.3612B}, and at $z \sim 8$ it has been shown to fit equally as well as a Schechter function \citep{2015ApJ...810...71F}. Further data over a wider area, such as the forthcoming BoRG[4JWST] survey, will be required to constrain the number density of the brightest galaxies in this epoch.

\section{Conclusions}
\label{sec:conc}

We have presented follow-up imaging of two bright high-$z$ galaxy candidates with the F098M filter on \emph{HST}/WFC3. Both galaxies were selected as $z \gtrsim 8$ candidates in the BoRG survey based on strong detections in the near-infrared and non-detections in the optical F350LP filter. The addition of another filter blueward of the break confirms BoRG\_0116+1425\_747 as a probable $z \sim 8$ source, but reveals that BoRG\_0116+1425\_630 - previously the brightest known $z > 8$ candidate - is likely to be a $z \sim 2$ interloper.

The removal of BoRG\_0116+1425\_630 from the $z > 8$ sample strongly affects the conclusions about the UV luminosity function in this epoch. Previously, the BoRG results of \citetalias{2016ApJ...817..120C} supported a transition from an exponential decline at the bright end after reionization to a shallower power-law decline beforehand. Removing this bright source from the sample means that there is now no evidence for a departure from the Schechter function, and therefore no evidence for difference in the galaxy formation process before and after reionization.

These results highlight the usefulness of the F098M filter for identifying $z < 2$ interlopers in Lyman Break-selected samples, and further demonstrates the need for large high-$z$ surveys to constrain the bright end of the luminosity function prior to the launch of \emph{JWST}.

\acknowledgments
The authors would like to thank the anonymous referee for helpful comments that clarified the text of this Letter. This work is based on observations made with the NASA/ESA Hubble Space Telescope, which is operated by the Association of Universities for Research in Astronomy, Inc., under NASA contract NAS 5-26555. We acknowledge support by NASA through grant HST-GO-14701. Parts of this research were supported by the Australian Research Council Centre of Excellence for All Sky Astrophysics in 3 Dimensions (ASTRO 3D), through project number CE170100013. R.C.L. acknowledges support from an Australian Research Council Discovery Early Career Researcher Award (DE180101240).

\bibliographystyle{aasjournal}
\bibliography{bib}

\end{document}